\documentstyle[prl,aps,epsf]{revtex}
\begin{document}
\draft
\twocolumn[\hsize\textwidth\columnwidth\hsize\csname @twocolumnfalse\endcsname
\title{Microscopic Viscoelasticity: Shear Moduli of Soft Materials \\ 
Determined from Thermal Fluctuations}
\author{F.~Gittes, B.~Schnurr, P.D.~Olmsted$^*$, 
F.C.~MacKintosh, and C.F.~Schmidt}
\address{Department of Physics \& Biophysics Research Division, \\ 
University of Michigan, Ann Arbor, MI 48109-1120 \\
$^*$Department of Physics, University of Leeds, Leeds LS2 9JT, \\ 
United Kingdom}
\maketitle
\begin{abstract}

We describe a high-resolution, high-bandwidth technique for
determining the local viscoelasticity of soft materials such as
polymer gels.  Loss and storage shear moduli are determined from the
power spectra of thermal fluctuations of embedded micron-sized probe
particles, observed with an interferometric microscope.  This provides
a passive, small-amplitude measurement of rheological properties over
a much broader frequency range than previously accessible to
microrheology.  We study both F-actin biopolymer solutions and
polyacrylamide (PAAm) gels, as model semiflexible and flexible
systems, respectively.  We observe high-frequency $\omega^{3/4}$
scaling of the shear modulus in F-actin solutions, in contrast to
$\omega^{1/2}$ scaling for PAAm.

\end{abstract}
\pacs{PACS numbers: 61.30.Cz, 61.30.-v, 64.70.Md, 83.70.Jr}
\vskip2pc]

The local analysis of viscoelasticity can explore the small-scale
structure of complex fluids.  It can also potentially measure bulk
viscoelastic quantities in small samples.  Experiments have been done
to characterize the local viscoelasticity of materials since at least
the 1920s, when magnetic particles in gelatin were manipulated by
field gradients \cite{Freundlich1922}.  Crick used a similar technique
to study living cells \cite{Crick1950}.  With the recent advent of
methods for force generation, detection, and manipulation of particles
on sub-micrometer scales, experimental possibilities have expanded
greatly, and interest in {\it microrheology\/} has grown substantially
\cite{microrheo,Ziemann94,Ziemann96,Amblard1996,UsMRS,Wirtz}.  Here we
report an optical technique for high resolution and high bandwidth
observations of the thermal fluctuations of particles embedded in soft
materials.  Using dispersion relations from linear response theory,
the frequency-dependent loss and elastic storage shear moduli are both
calculated from the fluctuation power spectra, which we measure with a
resolution of 2~\AA from 0.1~Hz to 20~kHz.  This frequency range exceeds
that of video-based microrheological experiments (although large-volume
rheometers can reach frequencies up to 500~kHz \cite{stokich}) and allowed us
to observe high-frequency $\omega^{3/4}$ scaling of the shear modulus for
entangled, semiflexible solutions. These dynamics differ
fundamentally from those of flexible polymer systems.

We have studied F-actin solutions as a model semiflexible polymer, and
polyacrylamide (PAAm) gels as a flexible polymer control.  Actin is
one of the primary components of the cytoskeleton of plant and animal
cells and is largely responsible for the viscoelastic response of
cells \cite{stossel}.  Actin is a particularly accessible model system
because individual filaments can be hundreds of microns in length.
Viscoelastic properties of entangled F-actin solutions {\em in vitro}
have been measured using conventional macroscopic rheology
\cite{pollard,janmeykas,kas,muller,macrorheo}. Actin has also been the
subject of recent microrheological studies
\cite{microrheo,Ziemann94,Ziemann96,Amblard1996}.

The relationship between thermal motion and hydrodynamic response is
well-known in the context of Brownian motion in simple viscous fluids.
Less obviously, the bulk viscoelastic properties of complex fluids can
also be determined from thermal motion.  Such a passive technique is
particularly useful for biopolymer systems which, on the one hand,
have a limited range of linear response, but which, on the other hand,
have a small enough shear modulus that the thermal motion of
micron-sized probes is detectable.  This idea was exploited recently
by Mason and Weitz \cite{Mason}, who used diffusing wave spectroscopy
(DWS) to observe average thermal fluctuations of concentrated
suspensions of probe particles and to measure rheological quantities
in complex fluids to high frequencies.  That technique measures light
multiply-scattered by many embedded particles, assumed to be
independent. While DWS requires large samples, the present technique
monitors isolated particles in very small samples.


For spherical probes, the response is given by a generalized Stokes
formula, which allows for quantitative determination of macroscopic
viscoelastic moduli.  In our experiments, rigid spherical beads of
radius $R$ move in a viscoelastic medium consisting of polymer plus
solvent. When a force $f$ acts on such a sphere, the deformation of
the surrounding medium can be calculated \cite{UsLater} to obtain the
linear response for the sphere: $x_{\omega} =\alpha(\omega)
f_{\omega}$, where $x$ is the position of the sphere. In general, the
response function is complex: $\alpha(\omega) =\alpha'(\omega)
+i\alpha''(\omega)$. Within certain limits described below, this
response function reflects pure shear motion of the medium and is
given by
\begin{equation}
\alpha(\omega)=\frac{1}{6\pi G(\omega) R}.
\label{eq:stokes}
\end{equation}
Here, the frequency-dependent complex shear modulus is given by
$G(\omega)= G'(\omega) +iG''(\omega)$, where $G'$ and $G''$ are the
storage and loss modulus.  (For the case of a purely viscous fluid
$G=-i\omega\eta$, leading to the well-known Stokes formula $f=6\pi\eta
R \dot x$).  The fluctuation-dissipation theorem \cite{landaustat}
relates $\alpha$, and thus $G$, to Brownian motions of the bead
induced by thermal fluctuations of the medium.

As in the case of a viscous fluid \cite{landaufluid}, Eq.\
(\ref{eq:stokes}) is valid provided that the viscous penetration depth
$\delta=\sqrt{2\eta/\rho\omega}$ is large compared with $R$, where
$\rho$ and $\eta$ are solvent density and viscosity; this is true for
frequencies up to about 1~MHz for a 1~$\mu$m bead in water.  The
validity of continuum viscoelasticity also requires that the bead is
at least larger than the mesh size of the network.

A somewhat more subtle limitation on all microrheological methods to
date, including ours, concerns the two-component nature of polymer
solutions. In a general viscoelastic medium, there can be another
relevant material parameter in addition to the shear modulus: the
compression modulus, or equivalently, the Poisson ratio.  In a simple
viscous fluid, only shear stress is relevant because of the
incompressibility of the fluid. In a polymer solution consisting of
both network and solvent, this is also true for the high-frequency
response because the network and incompressible solvent become
strongly coupled.  While a rigorous calculation of the response
requires a full treatment including both network and solvent aspects
of the material, we can estimate the validity of the approach given
above within a so-called two fluid model \cite{twofluid}, which takes
into account the viscous coupling of the two components.  We estimate
the frequency crossover from the balance between viscous and elastic
forces. The viscous force of the solvent on the network is $\eta
v/\xi^2$ per volume, where $\xi$ is the mesh size and $v$ is the
solvent velocity relative to the mesh. The local elastic force in the
network is $G \nabla^2 u \sim G u/R^2$ per volume at the bead surface,
where $u$ is the network displacement field. Thus viscous coupling is
dominant, and solvent and network move together, above a frequency
\begin{equation}
\omega_c \simeq G\xi^2/\eta R^2.
\label{eq:drainfreq}
\end{equation}

Above $\omega_c$, Eq.~(\ref{eq:stokes}) will hold. In our experiments
$\omega_c$ is of order 10~Hz ($G\simeq 1~{\rm Pa}$, $\xi/R\simeq 0.1$,
$\eta=1~{\rm cP}$), which is at the lower limit of our instrumental
resolution. At frequencies $\omega\ll\omega_c$, the response includes
compressional modes not present in Eq.~(\ref{eq:stokes}); these will
have wavelengths $q^{-1}\sim\xi(G/\eta\omega)^{1/2}$ and will make
$\alpha(\omega)$ independent of bead size.

Applying the fluctuation-dissipation theorem to the power spectral
density (PSD) of bead displacements $\langle x_\omega^2\rangle$ yields
\begin{equation}
\alpha''(\omega)=\frac{\omega}{2kT}\langle x^2_\omega\rangle .
\label{eq:fdt}
\end{equation}
When $\alpha''(\omega)$ is obtained over a large enough range of
frequencies, the complex response $\alpha(\omega)$ is obtained from
the Kramers-Kronig relations \cite{landaustat} by evaluating the
dispersion integral
\begin{equation}
\alpha'(\omega)=\frac{2}{\pi} P \int_0^\infty d\zeta
\frac{\zeta\alpha''(\zeta)} {\zeta^2-\omega^2}.
\label{eq:kk}
\end{equation}
The principal-value integral $P$ in Eq.~(\ref{eq:kk}) is conveniently
computed on a discrete data set as a sine transform of
$\alpha''(\omega)$ followed by a cosine transform. We finally take the
complex reciprocal of $\alpha(\omega)$ to obtain the complex
$G(\omega)$ by Eq.~(\ref{eq:stokes}). We have evaluated
Eq.~(\ref{eq:kk}) using model spectra with various upper and lower
cutoffs \cite{UsLater} and found it to be reliable over the scaling
range discussed below.

The shear modulus $G(\omega)$ typically exhibits three distinct
regimes in entangled flexible polymer solutions \cite{doi}. For a
non-crosslinked solution the behavior is essentially that of a viscous
liquid at frequencies lower than $1/\tau_R$, where $\tau_R$ is the
{\it reptation time\/}. Above $1/\tau_R$ a rubber-like plateau
appears, with a dominant, frequency-independent elastic response as in
a crosslinked gel. At even higher frequencies, above the relaxation
rate of the mesh\cite{doi}, the high-frequency moduli $G'$ and $G''$
are expected to obey a pure power-law, since no new characteristic
time appears until the molecular cutoff is reached.


Our samples were prepared as follows. Actin was extracted from chicken
skeletal muscle according to standard procedures \cite{pardee},
combined with silica beads, and polymerized (buffer: 2~mM HEPES
(pH~7.0), 2~mM MgCl$_2$, 50~mM KCl, 1~mM EGTA, 1~mM ATP) at a
concentration of 2~mg/ml under slow rotation to avoid bead
sedimentation. This yields a highly entangled solution of stiff
filaments with persistence lengths of order $10~\mu$m and a mesh size
$\xi \cong 0.2~\mu$m \cite{christoph}. PAAm gels were prepared as in
standard gel electrophoresis \cite{biorad} with concentrations of 2,
2.5, and 3\% weight/volume (20, 25, and 30~mg/ml); the PAAm contained
3\% bis-acrylamide as a crosslinker. These flexible polymers have a
persistence length on the order of nanometers; the mesh size of a 2\%
gel is about 50~$\rm\AA$ \cite{fawcett}. The smallest sample dimension
was their thickness of $70~\mu$m (F-actin) and $140~\mu$m (PAAm).

For bead position detection we couple a near-infrared laser into an
optical bench-mounted differential interference contrast (DIC)
microscope.  Polarized at $45^\circ$ to the axis of the DIC Wollaston
prism, the beam is split and the microscope objective produces two
overlapping foci ($\sim$200~nm apart) of orthogonal polarizations
\cite{interf}. A bead of higher refractive index than the medium in
the double focus causes a differential phase shift $\Delta\phi(x)$
between the beams which is a measure of bead position $x$.  After
recombination by a second Wollaston prism, the light, now elliptically
polarized, is passed through a quarter-wave plate and a polarizing
beam splitter onto two photodiodes. The normalized difference of the
photodiode currents is $\sin(\Delta\phi)$. Small excursions of a
centered bead result in a proportional voltage response which we
calibrated by controlled displacements of a piezoelectric stage. This
set-up is also used for optical trapping; here, the laser focus was
broadened and the power reduced ($\sim$0.6~mW in the specimen) so that
optical forces were negligible. Our resolution at high frequencies was
limited by shot noise to about $2\times 10^{-6}\rm~nm^2~Hz^{-1}$. The
bead-position signal was digitized at 60~kHz and filtered above 20~kHz
to eliminate aliasing.


In actin solutions, the power spectral density (PSD) of bead
fluctuations $\langle x_\omega^2\rangle$ was highly reproducible in
shape and amplitude, not only for different beads within a sample but
between samples and between different actin preparations
\cite{UsLater}. For samples with beads of diameter 0.9, 2.1, and
5.0~$\mu$m, the variation in PSD among six beads per sample (at
100~Hz) was $10-20$\%.

\begin{figure}[h]
\epsfxsize=\columnwidth
\centerline{\epsfbox{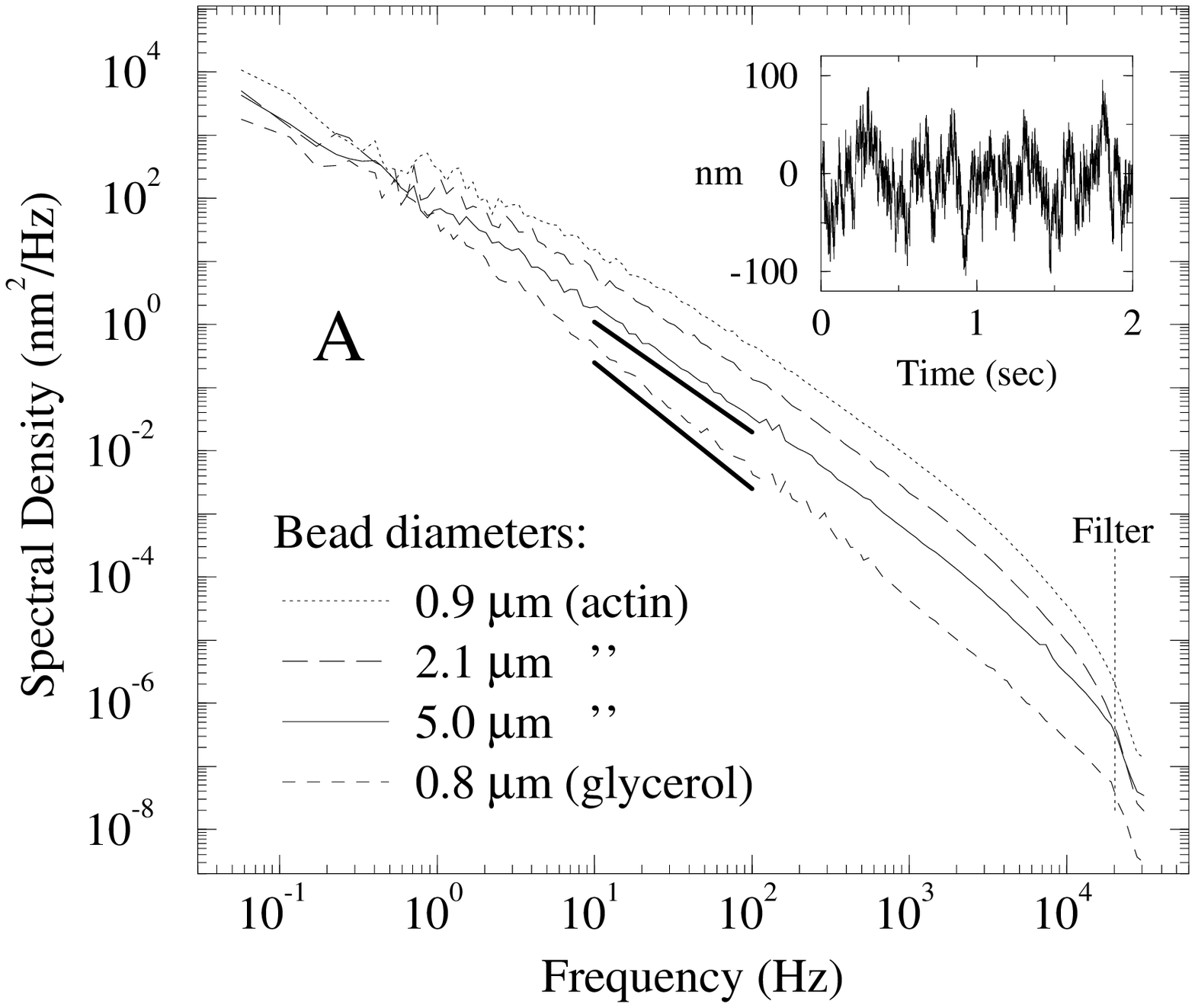}}
\epsfxsize=\columnwidth
\centerline{\epsfbox{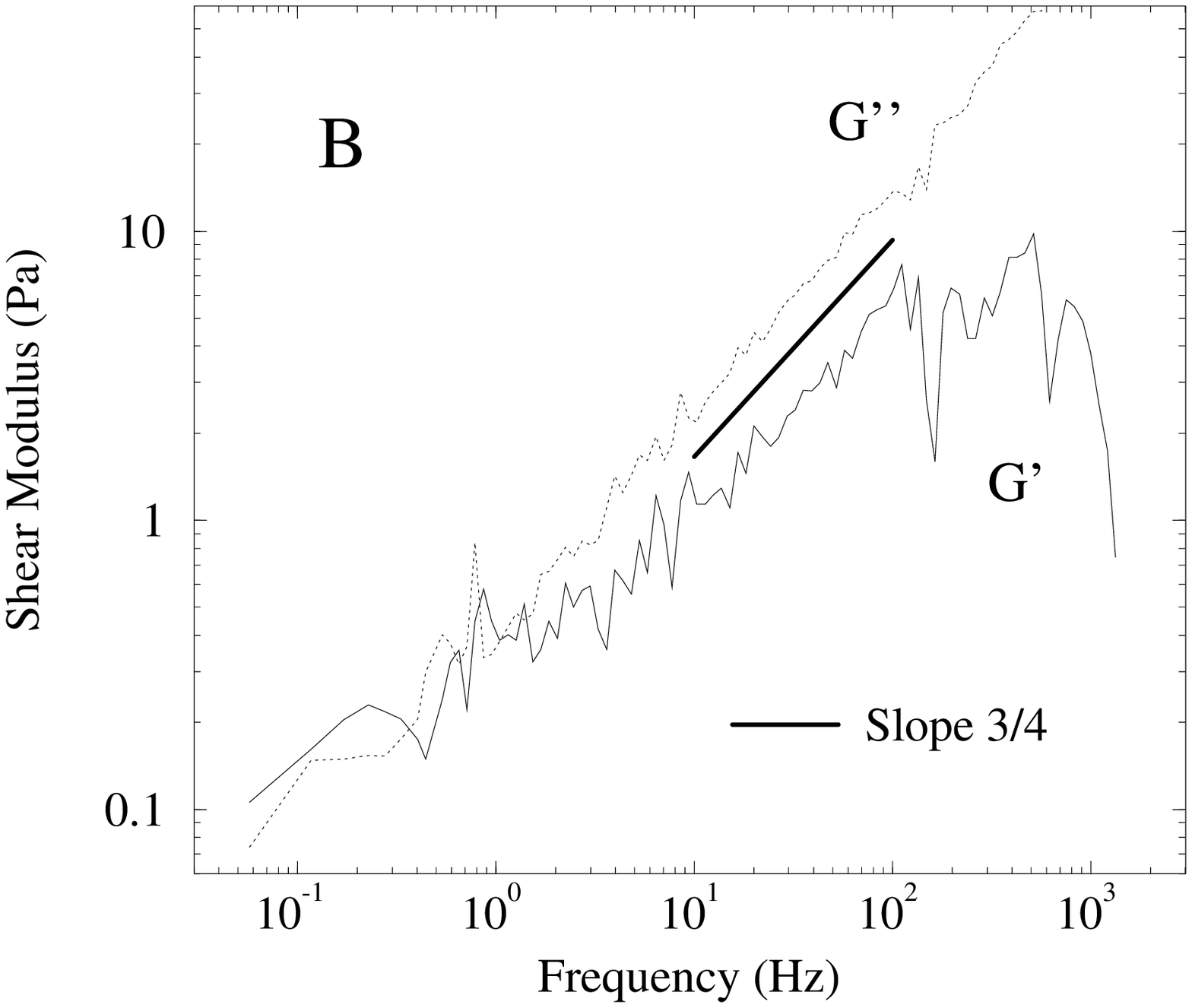}}
\caption{(A): Power spectral density (PSD) of thermal bead motion in an actin
solution (2~mg/ml), averaged over several beads and within bins of
equal log width. The PSD of a bead in glycerol is shown for
comparison.  The indicated slopes are $-1.75$ and $-2.0$.  Inset: Time
series data for a 0.9~$\mu$m bead.  For the smaller beads, deviations
from scaling may indicate a failure of the continuum assumption.  (B):
Real part (solid) and imaginary part (dotted) of the complex modulus
$G=G'(\omega)+iG''(\omega)$ calculated from the spectrum in (A)
(5.0~$\mu$m bead). $G'$ and $G''$ scale as $\omega^z$, with $z=0.76\pm
0.02$ determined from their ratio.  The downturn in the spectra at
about 20~kHz is due to an anti-aliasing filter.}
\label{fig:actin}
\end{figure}

The fluctuation power spectra of beads in actin are shown in
Fig.~\ref{fig:actin}.  Above about 20~Hz, the PSDs scale (roughly)
with bead size, as predicted by Eq.~(\ref{eq:stokes}). All spectra
show a slight downturn above 3~kHz, which is currently an experimental
limitation. At low frequencies, the slopes decrease in a bead-size
dependent way. We believe this to be the transition from a regime of
pure shear fluctuations to long-wavelength compressional modes, as
predicted by Eq.~(\ref{eq:drainfreq}). For 5.0~$\mu$m beads, the PSD
displays power-law behavior ($\sim\omega^{-n}$ with $n=1.77\pm0.02$)
over almost three decades of frequency.  In contrast, 0.8~$\mu$m
polystyrene beads in glycerol, a simple viscous fluid, should display
a diffusional exponent of $n=-2$; we find
$n=1.94\pm0.03$. Furthermore, from this spectrum we obtain a value of
11.3~P, compared to a tabulated value of 9.3~P at 25$^\circ$C.

The $G'(\omega)$ and $G''(\omega)$ calculated from the spectra for
5.0~$\mu$m beads are shown in Fig.~\ref{fig:actin}B. These beads are
about 20 times larger than the mesh size, so that the continuum model
of Eq.~(\ref{eq:stokes}) applies: as expected, a power-law has emerged
in $G'$ and $G''$. No low-frequency plateau is observed. (In
\cite{muller} this plateau was found below about 0.01~Hz at similar
concentrations.)  The scaling corresponds, via Eqs.~(\ref{eq:stokes})
and (\ref{eq:kk}), to a complex power-law $G(\omega)\propto
(i\omega)^{z}$, so that $G''/G'=\tan\pi z/2$. This ratio allows an
estimate of $z=n-1=0.76\pm0.02$ between 10~Hz and 100~Hz, consistent
with the power observed in the PSD.

Other studies of actin rheology \cite{Ziemann94,muller}, at
frequencies below about 3~Hz, have reported
$G\sim\omega^{1/2}$. However, at these lower frequencies, their
results may reflect an incomplete transition to the $\omega^{3/4}$
scaling regime. The magnitudes we measure for $G'$ and $G''$ are
consistent with other experiments
\cite{microrheo,Ziemann94,macrorheo}, although discrepancies persist
in the literature \cite{janmeykas}. Amblard {\it et al.\/}
\cite{Amblard1996} found bead-diffusion dynamics in more dilute actin
solutions consistent with our scaling results, but suggest an
explanation based on isolated filament dynamics. We believe that at
our higher concentrations, for which the mesh size is substantially
smaller than the bead diameter, a continuum elastic approach as
described here is correct.

\begin{figure}
\epsfxsize=\columnwidth
\centerline{\epsfbox{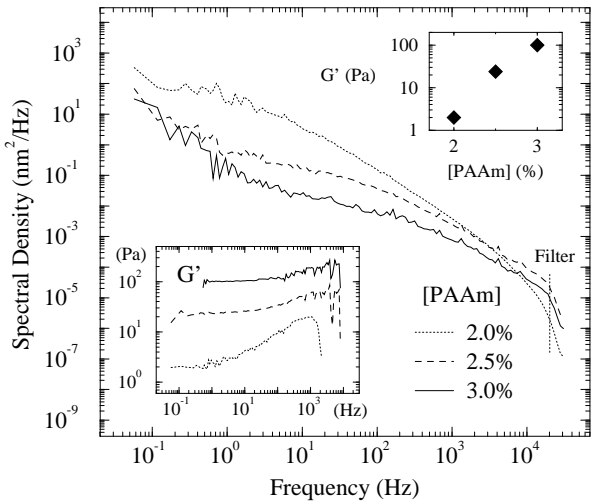}}
\caption{Power spectra of thermal motion of 0.9~$\mu$m beads in
polyacrylamide gels of different concentrations, averaged as in
Fig.~\protect{\ref{fig:actin}}A, cut off by an anti-aliasing filter at
20~kHz.  Lower inset: The elastic modulus $G'(\omega)$ obtained from each
of these spectra, as in Fig.~\protect{\ref{fig:actin}}B.  Upper inset:
Plateau moduli (values at 1~Hz), as a function of concentration, of the
three curves in the lower inset.}
\label{fig:paam}
\end{figure}

The fluctuation spectra of 0.9~$\mu$m silica beads in 2, 2.5, and 3\%
crosslinked PAAm gels are shown in Fig.~\ref{fig:paam}.  The mesh size
of this gel is about 50~$\rm\AA$ \cite{fawcett} so that a continuum
model easily applies. The 2\% spectrum has a slope of about $-1.5$
near 100~Hz, consistent with the Rouse model and in clear contrast to
the actin solutions. The 2\% spectrum decreases in slope below about
2~Hz due to the onset of an elastic plateau in $G'$ (the other spectra
reflect a plateau over most of their range, aside from the
low-frequency instrumental noise from drift below a few Hz).  The
calculated $G'(\omega)$ are shown explicitly in Fig.~\ref{fig:paam}
lower inset, where the plateau values of $G'$ are 2.0~Pa (2\%), 24~Pa
(2.5\%) and 100~Pa (3\%).  Because the plateau extends to such high
frequencies, $G'(\omega)$ does not yet exhibit the
$G(\omega)\sim\omega^{1/2}$ scaling \cite{doi} as suggested by the PSD
slope of $-1.5$.  Close to the gelation threshold, the shear modulus
increases very rapidly with concentration, as also found by others
\cite{nossal}.

The present technique differs in its spatial resolution and wide
frequency range from prior microrheological studies such as those of
Refs.~\cite{microrheo,Ziemann94,Ziemann96,Amblard1996} employing video
microscopy.  The extended frequency range has allowed us to determine
the high-frequency scaling of the shear modulus for actin well above
the elastic plateau.  We have described the inherent limitations in
this and similar microrheology studies at low frequencies. For rather
sparse networks, the low-frequency response cannot be described by a
shear modulus alone, but must also take into account the local
compression of the network; this may invalidate the generalized Stokes
formula, Eq.~(\ref{eq:stokes}), for frequencies accessible to video.
In comparison to most commercial rheometers, our technique allows the study of 
soft materials and small sample volumes, which may be particularly 
useful in biological applications.


This work was supported in part by the Whitaker Foundation, the
National Science Foundation (Grant Nos.\ BIR 95-12699 and DMR
92-57544), and by the donors of the Petroleum Research Fund,
administered by the ACS. We acknowledge generous technical support
from the Rowland Institute for Science, particularly W.\ Hill. We
thank P.\ Janmey, J.\ K\"as, and D.\ Weitz for helpful
discussions. PDO and FCM acknowledge support from NATO Grant No.\ CRG
960678.  FCM wishes to thank the Aspen Center for Physics.

\vspace{-1ex}


\end{document}